\documentstyle[prl,aps,twocolumn]{revtex}
\begin{document}
\draft

\title{	
\hfill {\small\rm SMC-PHYS-167}\\
Probabilistic naturalness measure for dipole moments due to new physics
}

\author{       Keiichi Akama,$^1$ Takashi Hattori$^2$ and Kazuo Katsuura$^3$}
\address{       $^{1,3}$ Department of Physics, Saitama Medical College,
                Moroyama, Saitama, Japan}
\address{       $^2$ Department of Physics, Kanagawa Dental College,
                Yokosuka, Kanagawa, Japan}
\maketitle
\begin{abstract}
We introduce a probabilistic measure of naturalness (naturalness level) 
	to fix naturalness bounds quantitatively.
It is applied to the anomalous magnetic moments and 
	the electric dipole moments due to new physics.
\end{abstract}



Quest for ``naturalness" \cite{naturalness_comp,naturalness_susy},
	without fine-tuning cancellations 
	motivates and promotes the people to develop
	beyond-the-standard model physics, such as 
	compositeness \cite{comp}, grand unification \cite{GUT}, 
	supersymmetry \cite{susy}, superstring \cite{superstring}, 
	braneworlds \cite{brane,brane_new}, and so on. 
The naturalness requirements, in addition to the experimental tests,
	serves as a guiding principle for new physics model building, 
	and bring us useful information on their parameters \cite{NB,Marciano,AHK}. 
In the previous paper \cite{AHK}, 
	we applied it to the naturalness of quark-lepton mass renormalization, 
	and derived many model-independent bounds 
	on their anomalous magnetic moments and electric dipole moments. 
The naturalness relations, however, have so far been
	ambiguous order-of-magnitude relations, 
	and have had no common basis to compare them model-to-model. 
In this paper, we would like to propose
	a universal probabilistic measure of naturalness,
	which we call ``naturalness level". 
We apply them to the previously derived naturalness bounds \cite{AHK}
	on the dipole moments.
For a given definite naturalness level,	
	we can assign more definite values for the bounds
	than the previous ones.

Let us consider what we usually mean by naturalness in physics.
For example, if the renormalized mass 
\begin{eqnarray}
	m^{\rm r}= m^{\rm b} +\delta m 
\label{m^r=}
\end{eqnarray}
	is very small compared with the bare mass $ m^{\rm b}$ 
	and its quantum correction $\delta m$,
	we say that it is unnatural fine-tuning cancellation. 
In saying so, we are expecting independent and random appearance 
	of $ m^{\rm b}$ and $\delta m$, 
	neglecting the further information at this stage.
If we know the true mechanism, it would no longer be unnatural.
In general, when we say something natural or unnatural,
	we seem always to assume some randomness 
	at some non-ultimate stage,  
	and if the probability based on the randomness is small 
	we say it unnatural.
Though the cancellations could have complicated form in general,
	the quantity should be defined by homogeneous function
	of its contributions 
	for meaningful comparison of their magnitudes. 
Thus we formulate it as follows.

Suppose, at some stage in some model, a quantity $X$ is given by 
\begin{eqnarray}
	X=f(X_1,X_2,\cdots),     	\label{X=}
\end{eqnarray}
	where $f$ is a homogeneous function 
	(i.e.\ $\forall a, aX=f(aX_1,aX_2,\cdots)$)
	of the contributions $ X_1,X_2,\cdots $. 
For example, for (\ref{m^r=}), we take 
	$X=m^{\rm r}$,  $X_1=m^{\rm b}$ and $X_2=\delta m$.
Then we assume that the $X_i$'s appear 
	with uniform randomness in $X_i$-space.
The naturalness means that 
	$|X|$ should not be too small
	in comparison with the individual contributions $X_i$,
	or, in general, with some homogeneous combination of them
	$ g(X_1,X_2,\cdots)$ 
	($\forall a, ag(X_1,X_2,\cdots)=g(aX_1,aX_2,\cdots)$).
Then, the naturalness bound is expressed as
\begin{eqnarray}
	|X|>\epsilon |g(X_1,X_2,\cdots)|  
\ \ \ \rm(naturalness\ bound)
\label{X>}
\end{eqnarray}
with some given small positive constant $\epsilon$.
For example, for (\ref{m^r=}), 
	$|m^{\rm r}|>\epsilon |\delta m|$, \ 
	$|m^{\rm r}|>\epsilon \sqrt{(m^{\rm b})^2+(\delta m)^2}$, etc..
Then we define the ``naturalness level (NL)" $p$ of the naturalness bound (\ref{X>})
	as its ``probability" based on the assumed randomness:
\begin{eqnarray}
	p=P(|X|>\epsilon |g(X_1,X_2,\cdots)|),  
\ \rm(naturalness\ level)
\label{p=}
\end{eqnarray}
which would be the statistical confidence level of the inequality (\ref{X>}),
	if the ``probability" be real probability.
Because we assume uniform randomness of $X_i$,  
	the naturalness level $p$ is the relative volume 
	occupied by the region of (\ref{X>}) in the $X_i$-space.
Since the functions $f$ and $g$ are homogeneous,  
	it is sufficient to evaluate the relative volume
	of the region (\ref{X>}) 
	on the hypersurface $\sum |X_i|^2=1$.

In the following, we derive the expressions 
	for the naturalness level $p$ in terms of $\epsilon$ 
	for the naturalness bounds in lepton- and quark-mass renormalizations.
We invert the relation to express $\epsilon$ in terms of $p$, 
	which enables us to fix the values of the naturalness bounds 
	for given naturalness levels.
Then, we apply them to examples of the naturalness bounds on 
	the anomalous magnetic moments and the electric dipole moments
	of leptons and quarks due to new physics \cite{AHK}.

\noindent
(A) {\bf Mass Renormalization under conserved CP}

In this case, the renormalized mass takes the form in (\ref{m^r=}) above.
Let us consider the naturalness bound 
\begin{eqnarray}
	|m^{\rm r}|>\epsilon |\delta m|.    \label{ m^r>}
\end{eqnarray}
In terms of the polar angle 
	$\phi=\tan^{-1}(X_2/X_1)= \tan^{-1}(\delta m/m^{\rm b})$  
	in the $X_1$-$X_2$ plane,
	(\ref{ m^r>}) is rewritten as
\begin{eqnarray}
	|\cos\phi+\sin\phi|>\epsilon |\sin\phi|.      \label{|cos...}
\end{eqnarray}
By the general arguments above,
	the naturalness level $p$ is $l/2\pi$ 
	with the total length $l$ of the arc 
	allowed by (\ref{|cos...}) on the unit circle 
	$ X_1^{\ 2} +X_2^{\ 2}=1$, and is given by
\begin{eqnarray}
	p ={1\over\pi}\cot^{-1}{\epsilon^2-2\over2\epsilon }.
\label{p=1}
\end{eqnarray}
Note that for $\epsilon\rightarrow 0(\infty)$, $p\rightarrow1(0)$.
For small $\epsilon$, 
\begin{eqnarray}
	p =1-{\epsilon \over\pi}-{\epsilon^3\over6\pi}
	+O(\epsilon^5).
\label{p=1-}
\end{eqnarray}
Eq.\ (\ref{p=1}) is inverted for $\epsilon$ as (with $q=1-p$)
\begin{eqnarray}
	\epsilon =\sqrt{2+\cot^2 p\pi}+\cot p \pi
	=\pi q-{\pi^3 q^3\over6}+O(q^5).  \label{eps=1}
\end{eqnarray}
This means that at NL (naturalness level) = 90\%(95\%),
\begin{eqnarray}
	|m^{\rm r}|>0.309 |\delta m|\ \ \ \ (|m^{\rm r}|>0.156 |\delta m|).  \label{m^r>1}
\end{eqnarray}

Now we apply this to the effects of the anomalous magnetic moment $\mu$ 
	due to new physics \cite{AHK}.
The contribution is given by
\begin{eqnarray}
	\delta m=-3eQ\mu\Lambda^2/8\pi^2,      \label{deltam}
\end{eqnarray}
where $e$ is the electric coupling constant, 
	$Q$ is the electric charge of the fermion,
	and $\Lambda$ is the new physics scale.
In practice, $\Lambda$ is the photon momentum cutoff 
	of the self mass diagram with the cutoff factor
	$(1-k^2/\Lambda^2)^{-2}$,
	where $k$ is the momentum of the photon propagator.
The lepton scattering experiments sets the bound  
\begin{eqnarray}
	\Lambda=(2\pi)^{-1/2}e\Lambda_-^{\rm VV}>2.26{\rm TeV},
      \label{LambdaQEDbound}
\end{eqnarray}
where we have used the phenomenological bound on
the contact-interaction scale $\Lambda_-^{\rm VV}$ = 18.0TeV
with vector-vector coupling and with negative sign \cite{Lambda_exp}.
(We use the value $e=0.315$ at the mass scale of 200GeV.) 

Using (\ref{LambdaQEDbound}) and the experimental values 
	for the masses \cite{PGD},
	we obtain the bounds for $\delta a=\mu/(e_0 Q/2m)$
	 ($e_0^2/4\pi=1/137$), at NL=90\%(95\%)
\begin{eqnarray}
&&	|\delta a_e|<9.1\times10^{-11}\ (1.8\times10^{-10}), \cr 
&&	|\delta a_\mu|<3.9\times10^{-6}\ (7.7\times10^{-6}), \cr 
&&	|\delta a_\tau|<1.1\times10^{-3}\ (2.2\times10^{-3}). \label{Adeltaa}
\end{eqnarray}
The subscripts of $\delta a$, $\mu$ and $d$ here and below
	indicate fermion species.

The recent experimental result on $\delta a_\mu$ \cite{Muon2}, 
	combined with the standard-model prediction \cite{mu_mu_th}, 
	set the 95\% confidence level lower bound 
\begin{eqnarray}
	\delta a_\mu>5.2\times10^{-10}.  \label{damuexp}
\end{eqnarray}
Combining (\ref{m^r>1}),( \ref{deltam}) and (\ref{damuexp}) we obtain the upper bound for the new physics scale $\Lambda$ at NL=90\%(95\%) 
\begin{eqnarray}
	\Lambda <\rm 200TeV\ \ \ \ (280TeV).\label{ALambda>}
\end{eqnarray}

For quarks, we use phenomenological values of the masses \cite{PGD}
	to obtain the bounds at NL=90\%(95\%)
\begin{eqnarray}
&&|\mu_{\rm u}|/\mu_{\rm N}<2.2\times10^{-6}\ (4.4\times10^{-6}), \cr 
&&|\mu _{\rm d}|/\mu_{\rm N}<8.4\times10^{-6}\ (1.7\times10^{-5}),\cr 
&&|\mu _{\rm s}|/\mu_{\rm N}<1.5\times10^{-4}\ (3.0\times10^{-4}), \cr 
&&|\mu _{\rm c}|/\mu_{\rm N}<6.9\times10^{-4}\ (1.4\times10^{-3}), \cr 
&&|\mu _{\rm b}|/\mu_{\rm N}<4.4\times10^{-3}\ (8.8\times10^{-3}), \cr 
&&|\mu _{\rm t}|/\mu_{\rm N}<9.1\times10^{-2}\ (1.8\times10^{-1}), 
\end{eqnarray} 
where $\mu_{\rm N}$ is the nuclear magneton.

For neutrinos, instead of (\ref{deltam}) we have
\begin{eqnarray}
	\delta m = - 3eg^2c\mu\Lambda^2/64\pi^4, \label{deltamnu}
\end{eqnarray}
where $g$ is the gauge coupling constant of 
	the electroweak gauge group SU(2)$\rm_L$, and
	$c$ is a numerical constant of $O(1)$,
	which is in principle calculable.
It requires, however, lengthy calculations and careful treatments
	of divergent integrals, and is presently under investigation. 
Apart from $c$, we obtain, at NL=90\%(95\%),
\begin{eqnarray}
&&	|\mu_{\nu_e}| c / \mu_{\rm B}<9.9\times10^{-14}\ (2.0\times10^{-13}), 
\cr &&	|\mu_{\nu_\mu}| c / \mu_{\rm B}<6.2\times10^{-9}\ (1.2\times10^{-8}), 
\cr &&	|\mu_{\nu_\tau}| c / \mu_{\rm B}<6.0\times10^{-7}\ (1.2\times10^{-6}), 
\label{Anu}
\end{eqnarray}
where $\mu_{\rm B}$ is the Bohr magneton.

\noindent
(B) {\bf Mass Renormalization with broken CP}
\\When CP invariance is broken, 
	the fermion mass have the 1- and $\gamma_5$- components
	(indicated by the subscripts 0 and 5, respectively),
\begin{eqnarray}
	m_0^{\rm r}= m_0^{\rm b} +\delta m_0,\ \ \   
	m_5^{\rm r}= m_5^{\rm b} +\delta m_5. 
\label{m^r=1}
\end{eqnarray}
The physical mass is given by the expression
\begin{eqnarray}
	m^{\rm r}=\sqrt{(m_0^{\rm b}+\delta m_0)^2+(m_5^{\rm b}+\delta m_5)^2}. 
\label{m =2}
\end{eqnarray}
which is taken as the $X$ in the above general arguments.  
It is homogeneous in $X_1=m_0^{\rm b}$, $X_2=\delta m_0$, 
	$X_3=m_5^{\rm b}$, and  $X_4=\delta m_5$,
	which are taken to be uniformly random.
Let us consider the naturalness bound
\begin{eqnarray}
	|m^{\rm r}|>\epsilon |\delta m_0|,
\ \ \ \   |m^{\rm r}|>\epsilon |\delta m_5|. \label{m^r>2}
\end{eqnarray}
The naturalness level $p$ is given by $V/2\pi^2$, 
	where $V$ is the volume of the part  
	allowed by (\ref{m^r>2}) on the hypersphere 
	$ \sum_{i=1}^4 |X_i|^{\ 2}=1$.
A lengthy but straightforward calculation leads to the expression
\begin{eqnarray}
	p=1-{2\over\pi}\int_0^{\phi_1}
	\sqrt{\alpha-\sin^2\phi\over\beta-\sin^2\phi }d\phi, 
\label{p=2}
\end{eqnarray}
where $\sin^2\phi_1=\alpha$ and
\begin{eqnarray}
&&	\alpha=
	\left[1-(1-\epsilon^2/2)/\sqrt{1+\epsilon^4/4}\right]/2,\ \cr
&&	\beta=
	\left[1+(1+\epsilon^2/2)/\sqrt{1+\epsilon^4/4}\right]/2. 
\label{alpha=}
\end{eqnarray}
For small $\epsilon$
\begin{eqnarray}
      p=1-{\epsilon^2\over8}\left(1+{3\epsilon^2\over16 }-{5\epsilon^4\over64 }
	+\cdots\right).
\label{p=eps}
\end{eqnarray}
It is inverted as (with $q=1-p$)
\begin{eqnarray}
      \epsilon=2\sqrt{2q}
	\left(1-{3q\over4 }+{143q^2\over32 }
	+\cdots\right).
\label{eps=}
\end{eqnarray}
This means that at NL 90\%(95\%),
\begin{eqnarray}
&&	|m^{\rm r}|>0.867 |\delta m_0|\ \ \ \ (|m^{\rm r}|>0.616 |\delta m_0|),  \cr
&&	|m^{\rm r}|>0.867 |\delta m_5|\ \ \ \ (|m^{\rm r}|>0.616 |\delta m_5|).  \label{m>2}
\end{eqnarray}
Note that, for small $p$, $p\sim O(\epsilon)$ in (A), 
	while $p\sim O(\epsilon^2)$ in (B),
	so that the corresponding bounds in (\ref{m>2})
	are more stringent than those in (\ref{m^r>1}).

Now we apply them to the effects of the anomalous magnetic moment $\mu$ 
	and the electric dipole moment $d$ due to new physics \cite{AHK}.
The contribution $\delta m_0$ is the same as the $\delta m$ in (\ref{deltam}), 
	while $\delta m_5$ is given by
\begin{eqnarray}
	\delta m_5=-3eQd\Lambda^2/8\pi.      \label{deltam5}
\end{eqnarray}
Then, we obtain at NL=90\%(95\%) 
\begin{eqnarray}
&&	|\delta a_e|<3.3\times10^{-11}\ (4.6\times10^{-11}), \cr
&&	|\delta a_\mu|<1.4\times10^{-6}\ (2.0\times10^{-6}), \cr 
&&	|\delta a_\tau|<3.9\times10^{-4}\ (5.5\times10^{-4}), \cr 
&&|d_e|<6.3\times10^{-22}\ (8.8\times10^{-22}) e{\rm cm}, \cr 
&&|d_\mu|<1.3\times10^{-19}\ (1.8\times10^{-19}) e{\rm cm}, \cr 
&&|d_\tau|<2.2\times10^{-18}\ (3.1\times10^{-18}) e{\rm cm}. \label{Bdeltaa}
\end{eqnarray}
The upper bound for the new physics scale $\Lambda$ 
	from (\ref{m^r>2}), (\ref{deltam}) and (\ref{damuexp}) 
	at NL=90\%(95\%) is
\begin{eqnarray}
	\Lambda <\rm 120TeV\ \ \ \ (140TeV). \label{BLambda>}
\end{eqnarray}
Note that bounds on $\delta a $'s in (\ref{Bdeltaa}) 
	and that on $\Lambda$ in (\ref{BLambda>}) are
	more stringent than those in (\ref{Adeltaa}) and (\ref{ALambda>}).
In general an additional condition (here the CP conservation) makes the bounds less stringent.

For quarks, we obtain the bounds at NL=90\%(95\%)
\begin{eqnarray}
&&|\mu_{\rm u}|/\mu_{\rm N}<7.9\times10^{-7}\ (1.1\times10^{-6}), \cr 
&&|\mu _{\rm d}|/\mu_{\rm N}<3.0\times10^{-6}\ (4.2\times10^{-6}),\cr 
&&|\mu _{\rm s}|/\mu_{\rm N}<5.4\times10^{-5}\ (7.7\times10^{-5}), \cr 
&&|\mu _{\rm c}|/\mu_{\rm N}<2.5\times10^{-4}\ (3.5\times10^{-4}), \cr 
&&|\mu _{\rm b}|/\mu_{\rm N}<1.6\times10^{-3}\ (2.2\times10^{-3}), \cr 
&&|\mu _{\rm t}|/\mu_{\rm N}<3.2\times10^{-2}\ (4.6\times10^{-2}), \cr 
&&|d_{\rm u}|<8.3\times10^{-21}\ (1.2\times10^{-20}) e{\rm cm}, \cr 
&&|d_{\rm d}|<3.1\times10^{-20}\ (4.4\times10^{-20}) e{\rm cm},\cr 
&&|d_{\rm s}|<5.7\times10^{-19}\ (8.1\times10^{-19}) e{\rm cm},\cr 
&&|d_{\rm c}|<2.6\times10^{-18}\ (3.6\times10^{-18}) e{\rm cm},\cr 
&&|d_{\rm b}|<1.7\times10^{-17}\ (2.3\times10^{-17}) e{\rm cm},\cr 
&&|d_{\rm t}|<3.4\times10^{-16}\ (4.8\times10^{-16}) e{\rm cm}.
\end{eqnarray}

For neutrinos, $\delta m_0$ is 
	the same as the $\delta m$ in (\ref{deltamnu}), 
	while $\delta m_5$ is given by
\begin{eqnarray}
	\delta m_5 = - 3eg^2cd\Lambda^2/64\pi^4. \label{deltam5nu}
\end{eqnarray}
Then, we obtain, at NL=90\%(95\%),
\begin{eqnarray}
&&|\mu_{\nu_e}| c / \mu_{\rm B}<3.5\times10^{-14}\ (5.0\times10^{-14}),\cr
&&|\mu_{\nu_\mu}| c / \mu_{\rm B}<2.2\times10^{-9}\ (3.1\times10^{-9}),\cr 
&&|\mu_{\nu_\tau}| c / \mu_{\rm B}<2.1\times10^{-7}\ (3.0\times10^{-7}),\cr 
&&|d_{\nu_e}| c<6.8\times10^{-25}\ (9.6\times10^{-25}) e{\rm cm}, \cr 
&&|d_{\nu_\mu}| c<4.3\times10^{-20}\ (6.1\times10^{-20}) e{\rm cm}, \cr 
&&|d_{\nu_\tau}| c<4.1\times10^{-18}\ (5.8\times10^{-18}) e{\rm cm}. \label{Bnu}
\end{eqnarray}

The presently available experimental bounds are
\begin{eqnarray}
&&	\delta a_e = (-1.2\pm2.8) 10^{-11}, \cite{exae}
\cr&&	d_e = (6.9\pm7.4) 10^{-28}e\rm cm, \cite{exde}
\cr&&	\delta a_\mu = (26\pm10) 10^{-11}, \cite{Muon2,mu_mu_th}
\cr&&	d_\mu = (3.7\pm3.4) 10^{-19}e\rm cm, \cite{exdmu}
\cr&&	-0.052<\delta a_\tau<0.058\ (95\%\rm CL), \cite{exatau}
\cr&&	-2.2<{\rm Re}d_\tau<4.5(10^{-17}e{\rm cm})\ (95\%\rm CL),\cite{exdtau}
\cr&&	-2.5<{\rm Im}d_\tau<0.8(10^{-17}e{\rm cm})\ (95\%\rm CL),\cite{exdtau}
\cr&&|\mu_{\nu_e}|<1.5\times10^{-10}\mu_{\rm B}\ (90\%\rm CL),\cite{exmunue}
\cr&&|\mu_{\nu_\mu}|<6.8\times10^{-10}\mu_{\rm B}\ (90\%\rm CL),\cite{exmunumu}
\cr&&|\mu_{\nu_\tau}|<3.9\times10^{-7}\mu_{\rm B}\ (90\%\rm CL),\cite{exmunutau}
\cr&&|d_{\nu_\tau}|<5.2\times10^{-17}e{\rm cm}\ (95\%\rm CL).\cite{exdnutau}
\end{eqnarray}
The naturalness bounds for $\delta a_\tau$, $d_\mu$, $d_\tau$, $\mu_{\nu_e}$,
	and $d_{\nu_\tau}$ are more stringent than the experimental bounds, 
	while those for $\delta a_e$, $\delta a_\mu$, $d_e$, 
	and $\mu_{\nu_\mu}$ are less stringent, 
	and that for $\mu_{\nu_\tau}$ is comparable.
If we use the information on the neutrino mass differences 
	from the experimental results on the solar \cite{solar} 
	and atmospheric \cite{atmo} neutrinos,
	the naturalness bounds for all the neutrinos become the same as
	those for $\nu_e$ in (\ref{Anu}) and (\ref{Bnu}),
	which are much more stringent than experimental ones.

We can apply the naturalness-level analysis to various problems.
For example, we can calculate the naturalness levels for  
	the mass ratios $m_e/m_\mu$, $m_\mu/m_\tau$, $m_u/m_t$, etc.\ 
	and the mixing angles and the CP violating phase,
	under the assumption that they are derived by the diagonalization
	of random mass matrices.
We can calculate the naturalness levels and compare them  
	under the various naturalness-achieving mechanisms,
	such as seesaw mechanism \cite{seesaw}, 
	compositeness \cite{naturalness_comp}, 
	supersymmetry \cite{naturalness_susy},
	braneworld etc \cite{brane_new}. 
Some of them are presently under investigations, 
	and will be reported elsewhere. 

The naturalness level defined here possesses very common quantitative meaning 
	independent of specific quantities in specific models.
It depends, however,  on the choice of the ``stage" 
	where we assume the randomness of the contributions to the quantity. 
The lower naturalness requires the further theoretical explanations
	from the more fundamental stage
	so as to improve the naturalness level.
We expect that such naturalness-level analyses
	would provide powerful quantitative guides 
	in search for true explanations of the nature.

\end{document}